\pdfoutput=1
\documentclass[%
 reprint,
 amsmath,amssymb,
 aps,
floatfix
]{revtex4-1}

\usepackage{graphicx}
\usepackage{dcolumn,color}
\usepackage{bm}
\usepackage{placeins} 

\begin{document}
\preprint{APS/123-QED}
\title{Leidenfrost temperature in sprays: role of the substrate and liquid properties}

\author{Fabian M. Tenzer}
\author{Julian Hofmann}
\author{Ilia V. Roisman}
\email{roisman@sla.tu-darmstadt.de}
\author{Cameron Tropea}
\affiliation{Institute for Fluid Mechanics and Aerodynamics, 
Technische Universit\"at Darmstadt\\
Alarich-Wei\ss-Stra\ss e 10, 64287 Darmstadt, Germany}
\date{\today}

\begin{abstract}
In this  study the Leidenfrost temperature during spray cooling of very hot substrates is experimentally measured. The spray parameters, i.e. the drop diameters and velocities and the mass flux, are very accurately measured. Astonishingly, the measured Leidenfrost temperature is independent of any of the spray impact parameters, but is determined exclusively by the materials of the liquid and the substrate.   

The mechanism of film boiling is explained by the formation of a fast propagating vaporizing front, when the inertial forces in the associated liquid flow are comparable with the viscous stresses. This leads to a theoretical prediction for the Leidenfrost temperature which agrees well with the experimental data. 

\end{abstract}

\maketitle

\paragraph*{Introduction} Spray impact onto very hot substrates is  associated with  spray cooling, spray lubrication or spray impingement on cylinder walls of  internal combustion engines. For example, sprays  are used  for cooling of hot forging tools \citep{Pola2013},  cooling of high power electronics \citep{mudawar2001} or for quenching of hot milled steel. The current state of the art regarding heat transfer during spray cooling is well summarized in a number of recent comprehensive review articles \citep{Liang2017b,Liang2017d,Cheng2016a,Kim2007,Breitenbach2018}. 

Various thermodynamic and hydrodynamic regimes have been identified during contact of liquid with a hot substrate, including conduction, nucleate boiling, transition boiling, thermal atomization and film boiling. The physics of the transition from the nucleate boiling regime  to the film boiling regime at the Leidenfrost point is not yet completely known in that the Leidenfrost temperature cannot be reliably predicted. Several theoretical models have been developed based on the hydrodynamic stability analysis of the vapor/liquid interface \cite{Jerome1960,Zuber1958,Kakac2008} or thermocapillary stability \cite{Aursand2018}. Some authors assume that the Leidenfrost temperature is determined by the foam limit \cite{Spiegler1963,Wang2019}, which is the maximum temperature to which a liquid can be superheated, or by the limiting minimum vapor thickness \cite{Cai2020} when it becomes comparable with the surface roughness. However, the influence of the surface roughness is not yet clearly delineated and requires further investigations.

Numerous studies deal with the Leidenfrost point during \textit{single drop impact} \citep{Quere2013, Biance2003,Castanet2015,Tran2012}, showing a difference between the static Leidenfrost temperature $T_\mathrm{Ls}$ of a sessile drop and the dynamic Leidenfrost temperature $T_\mathrm{Ld}$ of an impacting drop. 
In these studies the dynamic Leidenfrost temperature is often identified as the lower bound for dry rebound of the drop  \cite{Bernardin1999}. 

There is an ongoing discussion about the effect of the impact parameters on the Leidenfrost point \citep{Liang2017c}. In many studies the Weber number $We = \rho_\mathrm{f} D U^2/\sigma$
is considered  an important parameter \citep{Bertola2015a,Yao1988}, where $D$, $U$, $\rho_\mathrm{f}$ and $\sigma$ are the mean drop diameter, mean drop velocity, density and surface tension. However, recently it was experimentally demonstrated that the Weber number is not a relevant influencing number for the thermodynamic phenomena associated with boiling \cite{Roisman2018}. 

The \textit{Leidenfrost point during spray impact} is usually defined as the temperature at which the heat flux reaches a minimum.  This definition has been recently confirmed by high-speed video observations \cite{Tenzer2019}. 

Several studies on spray impact onto a hot substrate \cite{Gottfried1966b,Hoogendoorn1974,Al-Ahmadi2008, Guven2010, Sozbir2003} indicate an increase of the Leidenfrost temperature for higher mass flux $\dot{m}$. In  dimensionless form the effect of the mass flux is often described as a dimensionless spray Weber number $We_\mathrm{S}=\dot{m}^2 D/\rho_\mathrm{f} \sigma$  \cite{Yao2002,Labergue2015}. 
The correlations presently available for predicting the Leidenfrost point in a spray are  completely empirical. The spray Weber number $We_\mathrm{S}$ \cite{Labergue2015},  the average drop velocity \cite{Klinzing1992a,Bernardin1996a,Bernardin2004b} or mass flux \cite{Al-Ahmadi2008} are among the influencing parameters in these correlations. 

In the present study the Leidenfrost point is experimentally determined using accurately  characterized sprays whose impact parameters are varied over wide ranges. A theory for predicting the Leidenfrost point is then introduced, yielding excellent agreement with measurements

\begin{figure}
	\centering\includegraphics[width=0.9\linewidth]{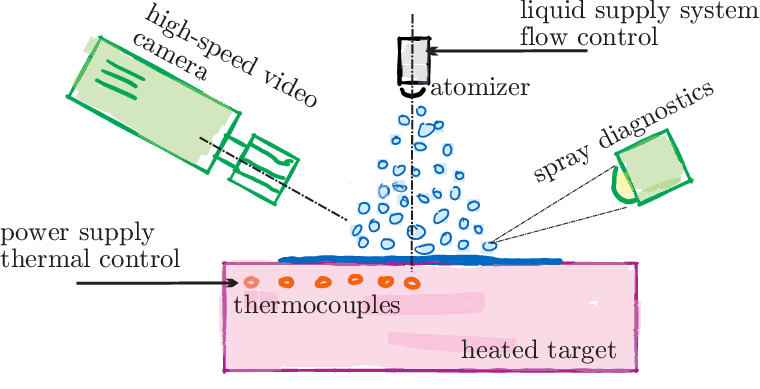}
	\caption{Sketch of the experimental setup.}
	\label{fig:setup}
\end{figure}

\paragraph*{Experimental method and measurement results} 
The experimental setup, as shown in Fig.~\ref{fig:setup}, consists of a heated target, a spraying system, a high-speed visual observation system and spray characterization system. 
The temperature distribution in the target, heated using cartridge heaters,  is measured by a set of the thermocouples placed in two rows at different depths from the surface. These temperatures are used for computation of the local heat flux and instantaneous surface temperature by solution of the inverse heat conduction problem \cite{Woodfield2006}. 

To determine the effect of the wall thermal properties on the Leidenfrost point two targets of different materials, stainless steel and nickel, have been used for the experiments. 

The mean drop diameters and velocities of the sprays generated by the atomizers  were measured using a phase Doppler instrument. For measurements of the local mass flux density  a custom built patternator was used. The properties of the target material and more detailed description of the experimental systems can be found in the supplementary material \cite{suppl1}.

\begin{figure}
	\centering\includegraphics[width=0.9\linewidth]{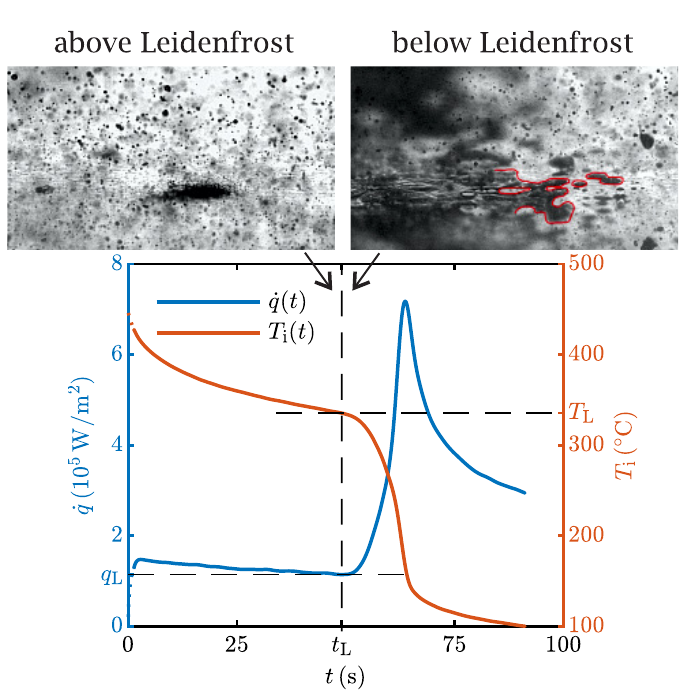}
	\caption{Exemplary results for the evolution of the heat flux $\dot{q}$ and surface temperature $T_\mathrm{i}$ as a function of time $t$ for spray cooling with distilled water. Leidenfrost temperature $T_\mathrm{L}$, heat flux at  the Leidenfrost point $\dot{q}_\mathrm{L}$ and instant of the Leidenfrost point $t_\mathrm{L}$ (both indicated by dashes lines) are chosen where the heat flux reaches its minimum $\dot{q}_\mathrm{L}$. Inserts show  liquid patterns on the surface shortly above and below the Leidenfrost point. A liquid patch on the surface is outlined in red.}
	\label{fig:q_T_min}
\end{figure}

Typical results of the measurements of the heat flux $\dot{q}(t)$ and the wall interface temperature $T_\mathrm{i}(t)$ as functions of time $t$ are shown in Fig.~\ref{fig:q_T_min} for a stainless steel target, initially heated uniformly up to $T_\mathrm{w}\approx 450\, \mathrm{^\circ C}$. At some instant, $t=0$, the heating is switched off and the spray is simultaneously switched on. The example results in this figure are taken from one particular position on the target.

Due to the continuous cooling, the surface temperature $T_\mathrm{i}(t)$  decreases monotonically with time. The minimum of the heat flux curve $\dot q(t)$ determines the Leidenfrost point ($q_\mathrm{L}$, $T_\mathrm{L}$ and $t_\mathrm{L}$) and the maximum corresponds to the critical heat flux. 

The inserts in Fig.~\ref{fig:q_T_min}  compare the flow patterns on the  surface just below and above the Leidenfrost point. At temperatures higher than the Leidenfrost temperature (the left image), there are no remaining wet patches after drop impact and their rebound. At  wall temperatures slightly below (the right image), first very small patches of liquid remain on the surface following  drop impact. Evaporation of these patches leads to the rapid increase of  heat flux at $t>t_\mathrm{L}$ during the short transitional boiling regime, until the critical heat flux is reached.

\paragraph*{Effect of  spray parameters on the Leidenfrost temperature} The difficulties in analysis of spray impact and spray cooling phenomena are caused by the fact that it is not easy to vary spray parameters - mass flux, average drop diameter and magnitude of the drop velocity - independently. On the other hand, the effect of the mass flux can only be significant  if the probability of  multiple drop interactions at the  surface is high. The measure of this probability  is related to the relative wetted area $\lambda$ of the wall surface for each drop impact. For high  Weber numbers this parameter can be expressed in the form $\lambda \sim \dot q We^{0.96}/\rho_\mathrm{f} U$ in the film boiling regime \cite{Breitenbach2017}. 
In many practical cases and in all of the present experiments the value of the relative wetted area $\lambda$ is much smaller than unity. The spray  can therefore be considered simply as the superposition of single drop impacts. This conclusion is supported by the heat flux measurements in the film boiling regime \cite{Tenzer2019}.

\begin{figure}
	\centering\includegraphics[width=.8\linewidth]{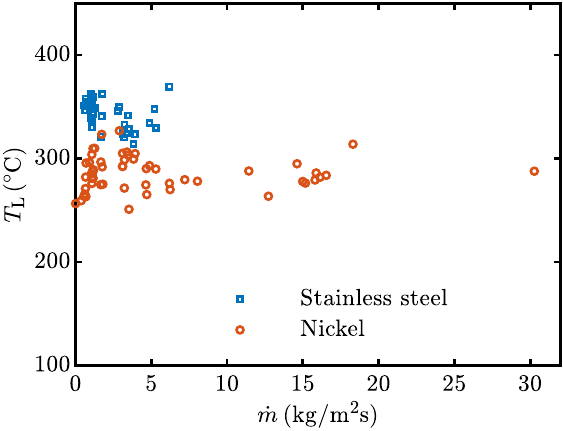}
	\caption{Leidenfrost temperature $T_\mathrm{L}$ as a function of the mass flux $\dot{m}$.}
	\label{fig:TL_m}
\end{figure}

The dependence of the Leidenfrost temperature on the mass flux of the impinging spray is shown in Fig.~\ref{fig:TL_m} for stainless steel and nickel targets. As expected, there is no clear correlation between the Leidenfrost temperature and the mass flux of the impinging spray for both materials. 

The data for stainless steel is shown only for the mass fluxes $\dot q < 7$ kg/m$^2$s. For higher mass fluxes the heat flux is rather high and the evolution of the interface temperature is extremely fast. Under these conditions the duration of the film boiling regime is comparable to the rise time of the thermocouples. Therefore, only  data below this limit are shown in Fig.~\ref{fig:TL_m}  for which the  interface temperature can be accurately resolved and thus the Leidenfrost temperature can be correctly determined. 

While no influence of the mass flux can be recognized, the effect of the target material on the Leidenfrost temperature is apparent and significant.

\begin{figure}
	\centering\includegraphics[width=0.49\linewidth]{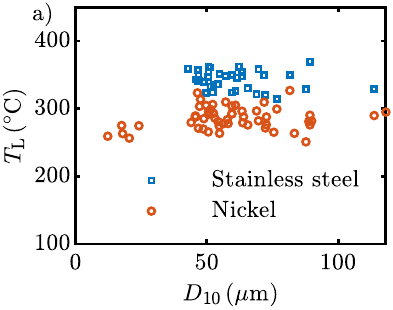} \includegraphics[width=0.49\linewidth]{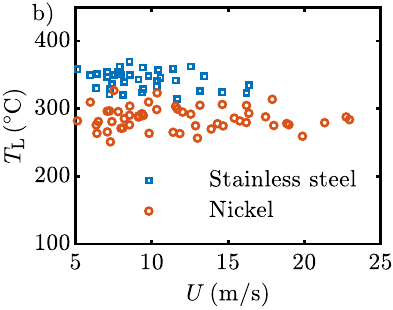}
	\\
	\centering\includegraphics[width=0.49\linewidth]{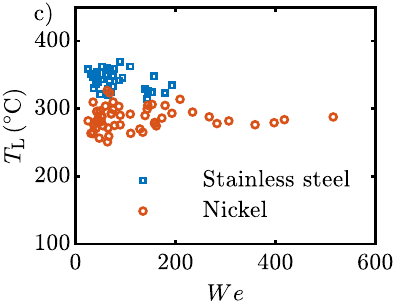} \includegraphics[width=0.49\linewidth]{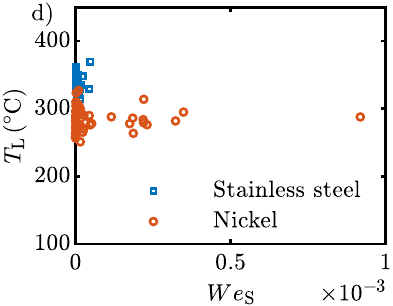}	
	\caption{Dependence of the Leidenfrost temperature $T_\mathrm{L}$ on a) mean drop diameter $D_\mathrm{10}$, b) mean drop velocity $U$, c) Weber number $We=\rho_\mathrm{f} D_{10} U^2/\sigma$ and d) spray Weber number $We_\mathrm{S}=\dot{m}^2 D_{10}/\rho_\mathrm{f} \sigma$.}
	\label{fig:T_L_par}
\end{figure}

It is interesting to examine how other mean or integral spray parameters influence the value of the Leidenfrost temperature. In Fig.~\ref{fig:T_L_par} the dependence of $T_\mathrm{L}$ on the mean drop diameter $D_{10}$, mean drop velocity $U$, the Weber numbers $We$ and on the spray Weber number $We_\mathrm{S}$ are shown for sprays cooling stainless steel and nickel targets. No correlation between $T_\mathrm{L}$ and any of the considered spray parameters can be identified. This is a rather surprising result which can significantly simplify modeling of transient cooling of hot surfaces.

\begin{figure}
    \centering
    \includegraphics[width=.8\linewidth]{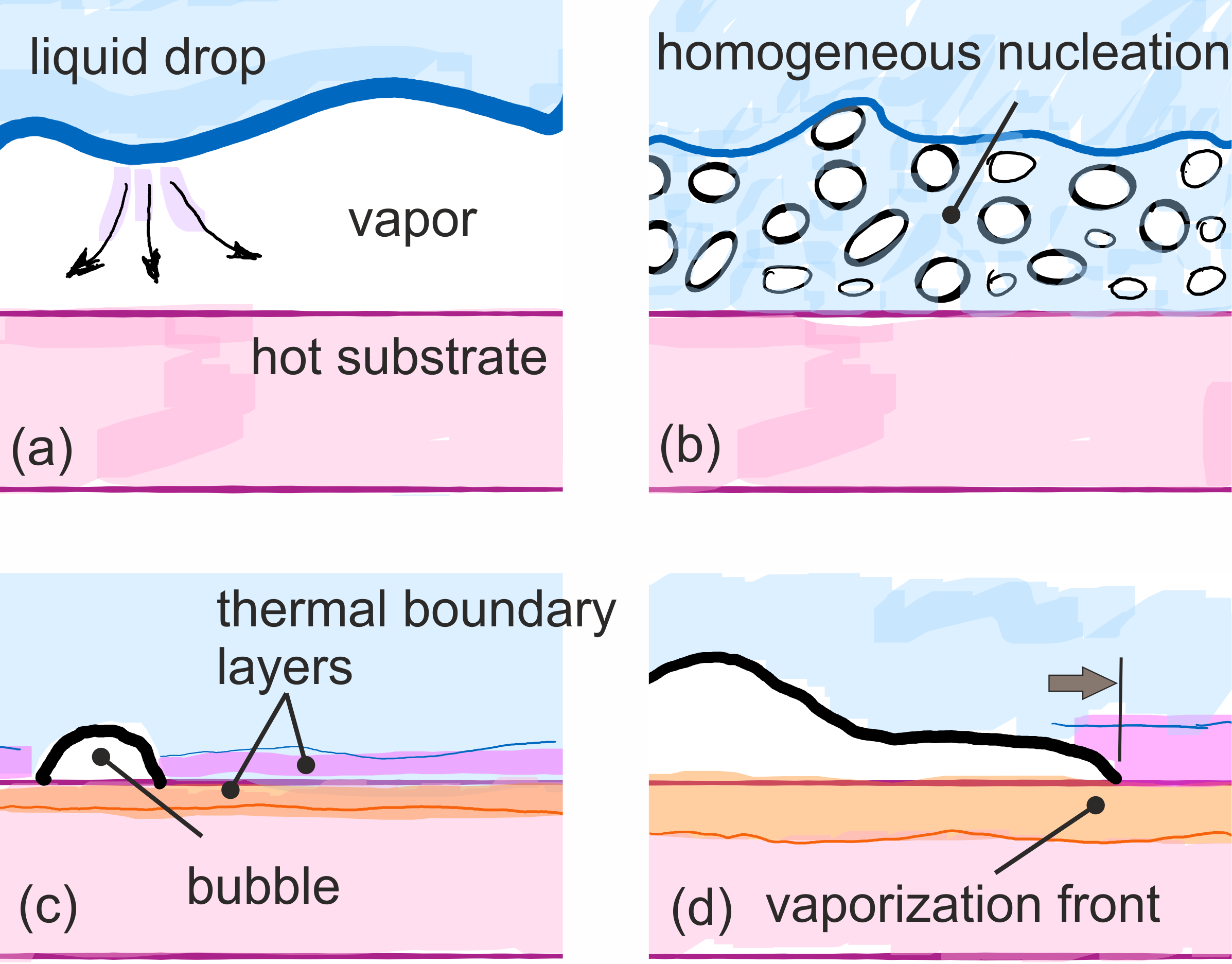}
    \caption{Possible mechanisms of film boiling: (a) stabilization of  a liquid-vapor interface by the vapor generation and (b) homogeneous nucleation at a certain temperature. A single bubble (c) can also expand along the substrate due to strong vaporization at the contact line (d). }
    \label{fig:mechanisms}
\end{figure}

The central question is \textit{What happens at the Leidenfrost point?} This question is first pursued by examining possible mechanisms for stabilizing the vapor-liquid interface occurring in the film boiling condition. One mechanism is the enhancement of  vaporization in the thinner regions of the vapor layer due to the higher  heat flux there, as shown schematically in Fig.~\ref{fig:mechanisms}a. Such a mechanism counteracts the  Rayleigh-Taylor instability.  Thus, the transition to the nucleate boiling regime is not possible, since the heat flux, and thus the evaporation rate, become infinite as soon as the vapor layer thickness goes to zero. 

Another potential reason for instability of the film boiling regime could be the homogeneous nucleation which is initiated in the liquid at a certain temperature \cite{Avedisian1985}, as illustrated in Fig.~\ref{fig:mechanisms}b. This assumption has been used in  models for the Leidenfrost temperature \cite{Wang2019,Bjornard1977,Wang2019a}, in particular, for the explanation of the dependence of its value on the thermal effusivity of the target material. The estimated homogeneous nucleation temperature is  $T_t=306^\circ$ C,  computed from the data for the Leidenfrost temperature \cite{Wang2019a}. Another estimation \cite{Avedisian1985} for the homogeneous nucleation temperature is  $T_t=202^\circ$ C. 
However this assumption contradicts  the observations of  nucleate boiling on  substrates with very high thermal effusivity, where the Leidenfrost point approaches the saturation temperature. The measured Leidenfrost temperature for water drops on  copper targets is 124$^\circ$ C for a sessile drop or 134$^\circ$ C for impacting drops in a spray, which are both lower than the  limit for superheat for distilled water. Moreover, neither of these hypotheses can explain the observed significant  influence of nano-structures, surface morphology or wettability of the substrate on the  Leidenfrost temperature  \cite{Auliano2016,Kruse2013,Romashevskiy2016,Kim2012,Takata2005}. 

Consider however the first stage of heat transfer, just after the contact of the liquid with the solid substrate. In the nucleate boiling regime the contact is followed by the emergence of numerous bubbles on the solid substrate as a result of heterogeneous nucleation. The heat transfer can be described by the heat conduction in two expanding thermal boundary layers in the liquid and the substrate respectively. The thickness of the thermal boundary layer is $h\sim\sqrt{\alpha t}$ where $\alpha$ is the thermal diffusivity of the corresponding material, liquid or solid. This phenomenon is depicted  schematically in Fig.~\ref{fig:mechanisms}c. 

The typical heat flux at the solid-liquid interface in the nucleate boiling regime is determined by the thickness of the boundary layer in the solid substrate and by the fact that the temperature of the vapor bubble interface is close to the saturation temperature $T_\mathrm{sat}$
\begin{equation}\label{eq:qsat}
    \dot q \sim\frac{e_\mathrm{w} (T_\mathrm{w0}-T_\mathrm{sat})}{\sqrt{t}},
\end{equation}
where $T_\mathrm{w0}$ is the initial substrate temperature prior to contact with the liquid and $e_\mathrm{w}$ is the thermal effusivity of the wall material. Expression (\ref{eq:qsat}) has been recently validated by comparison with the experimental data for single drops \cite{Breitenbach2017b} and sprays \cite{Tenzer2019}. 

At some elevated wall temperature, characterized by intensive heating,  vaporization fronts appear instead of local bubbles and expand along the substrate with a certain velocity. For a high enough velocity the front leads to the formation of a thin near-wall vapor layer (as shown in the sketch in Fig.~\ref{fig:mechanisms}d).  The propagation velocity of the vaporization front, depending on the heating rate and liquid properties, can be constant \cite{Okuyama2006,Avksentyuk2000,Stutz2013,Aktershev2011} or can grow exponentially \cite{Staszel2018} in time. 

We can roughly estimate the characteristic velocity $U_\mathrm{vap}\sim \dot q/\rho_\mathrm{f} L$ of the vaporization front using (\ref{eq:qsat})
\begin{equation}\label{eq:vapvel}
    U_\mathrm{vap}\sim \frac{e_\mathrm{w} ( T_\mathrm{w0}-T_\mathrm{sat})}{\rho_\mathrm{f} L \sqrt{t}},
\end{equation}
where $L$ is the latent heat of evaporation and $\rho_\mathrm{f}$ is the  density of the liquid. A similar approach has been successfully used to predict the velocities of  secondary drops in the thermal atomization regime   \cite{Breitenbach2018a,Roisman2018}. 

The only available characteristic thickness in the liquid region associated with the vaporization front is the thickness of the thermal boundary layer in the fluid $h_\mathrm{vap}\sim\sqrt{\alpha_\mathrm{f} t}$. Therefore, the Weber and the capillary numbers based on the typical vaporization front velocity (\ref{eq:vapvel}) are singular at $t\rightarrow 0$ which means the influence of the surface tension on the formation of the vaporization front is initially negligibly small. The Reynolds number $Re_\mathrm{vap}=\rho_\mathrm{f} h_\mathrm{vap} U_\mathrm{vap}/\mu_\mathrm{f}$, in contrast, is finite
\begin{equation}\label{eq:r}
   Re_\mathrm{vap}=\frac{e_\mathrm{w} \alpha_\mathrm{f}^{1/2} ( T_\mathrm{w0}-T_\mathrm{sat})}{\mu_\mathrm{f} L}.
\end{equation}
The viscous stresses therefore play an important role during the entire process of the vaporization front formation. The characteristic superheat $\Delta T^*$ at which the inertia and viscous terms are comparable can be determined assuming $Re_\mathrm{vap}=1$ in (\ref{eq:r})
\begin{equation}
    \Delta T^*=\frac{\mu_\mathrm{f} L}
{e_\mathrm{w} \alpha_\mathrm{f}^{1/2}}
\end{equation}

\begin{figure}
	\centering\includegraphics[width=.8\linewidth]{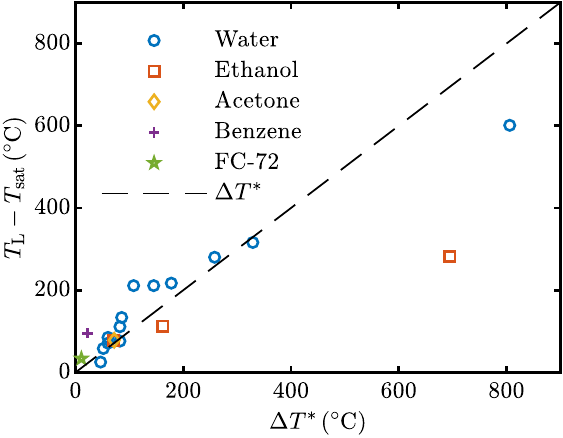}
	\caption{Substrate superheat at the Leidenfrost point, $T_\mathrm{L}-T_\mathrm{sat}$, as a function of $\Delta T^*$ for sessile drops. The data for different liquids, from this study and from \cite{Baumeister1973,Bernardin1999,Emmerson1975,Wachters1966,Wang2019a}, are listed in the Supplementary Material \cite{suppl1}.}
	\label{fig:T_Lm_e}
\end{figure}

\begin{figure}[b]
	\centering\includegraphics[width=.8\linewidth]{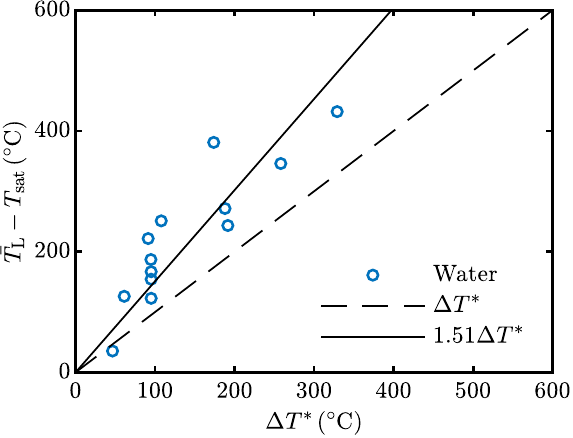}
	\caption{Mean substrate superheat at the Leidenfrost point $\overline T_\mathrm{L}-T_\mathrm{sat}$ as a function of $\Delta T^*$  for spray cooling. The data  for distilled water  from this study and from \cite{Hoogendoorn1974,Shoji1984,Ito1991,Yao1987,Choi1987,Yao2002,Bernardin1997b,Tran2012,Wang2019a} are listed in the Supplementary Material \cite{suppl1}.}	\label{fig:T_Lm_spray}
\end{figure}

In Fig.~\ref{fig:T_Lm_e} the experimental data for $T_\mathrm{L}-T_\mathrm{sat}$ of sessile drops of different liquids on various substrates is shown as a function of $\Delta T^*$. The liquid properties for calculation of $\Delta T^*$ are taken at the saturation temperature. It seems that for a wide range of targets the value of $T_\mathrm{L}-T_\mathrm{sat}$ can be predicted rather well as  $T_\mathrm{L}-T_\mathrm{sat}=\Delta T^*$. For  wall material with the smallest thermal effusivity the values for the Leidenfrost temperature deviate significantly from the predictions. In both cases, for water and for ethanol, the measured wall temperatures are above the critical temperature. In this case  additional physics has to be accounted for.

The corresponding data for spray impact or high Weber number drop impact are shown in Fig.~\ref{fig:T_Lm_spray}. The superheat associated with the Leidenfrost temperature also follows  a linear dependence on $\Delta T^*$. The best fit of the data is $\overline T_\mathrm{L}-T_\mathrm{sat}=1.51\Delta T^*$. The Leidenfrost temperature in the impacting drops increases, since the thickness of the thermal boundary in the liquid \cite{ROISMAN2010} and the propagation of the vaporization front are influenced by the flow in the spreading drop. 

The fact that the value of $T_\mathrm{L}-T_\mathrm{sat}$ is comparable with $\Delta T^*$ for a range of different liquids and substrates can indicate that the  initiation of the  film boiling can  indeed be explained by the propagation of the vaporization front along the substrate, assuming that the wall temperature does not exceed the critical value for the studied liquid. Note that propagation of the vaporization front is also influenced by the conditions at the moving contact line. This explains  the  sensitivity of the Leidenfrost temperature  to the substrate micro- and nano-morphology and wettability, identified in the existing literature.    

The authors gratefully acknowledge financial support from the Deutsche Forschungsgemeinschaft (DFG) in the framework of SFB-TRR 75 and the Industrieverband Massivumformung e.V.

\FloatBarrier
\bibliographystyle{apsrev4-2}
\bibliography{library}

\onecolumngrid
\clearpage

\begin{center}
	\textbf{\large Supplemental Material: Leidenfrost temperature in sprays}
\end{center}

\FloatBarrier

\section{Details on the experimental method} The experimental setup, as shown in Fig.~\ref{fig:setup}, consists of a heated target, a spraying system, a high-speed visual observation system and a spray characterization system.

The target ($100\, \mathrm{mm}$ diameter, $53.2\, \mathrm{mm}$ height) is equipped with 15 thermocouples (type K, $0.5\, \mathrm{mm}$ shield diameter, open measurement tip) and heated from the bottom by cartridge heaters. The thermocouples are placed in two rows (12 elements $0.5\, \mathrm{mm}$ and 2 elements $20\, \mathrm{mm}$ below the surface). Two targets of different materials have been used for the experiments to determine the effect of the wall thermal properties on the Leidenfrost point. One target is made from  stainless steel (1.4841) and  the other  from nickel (2.4068). In both cases the  surface exposed to spray impact is mirror polished. 

\begin{figure}
	\centering
	\includegraphics[width=0.5\linewidth]{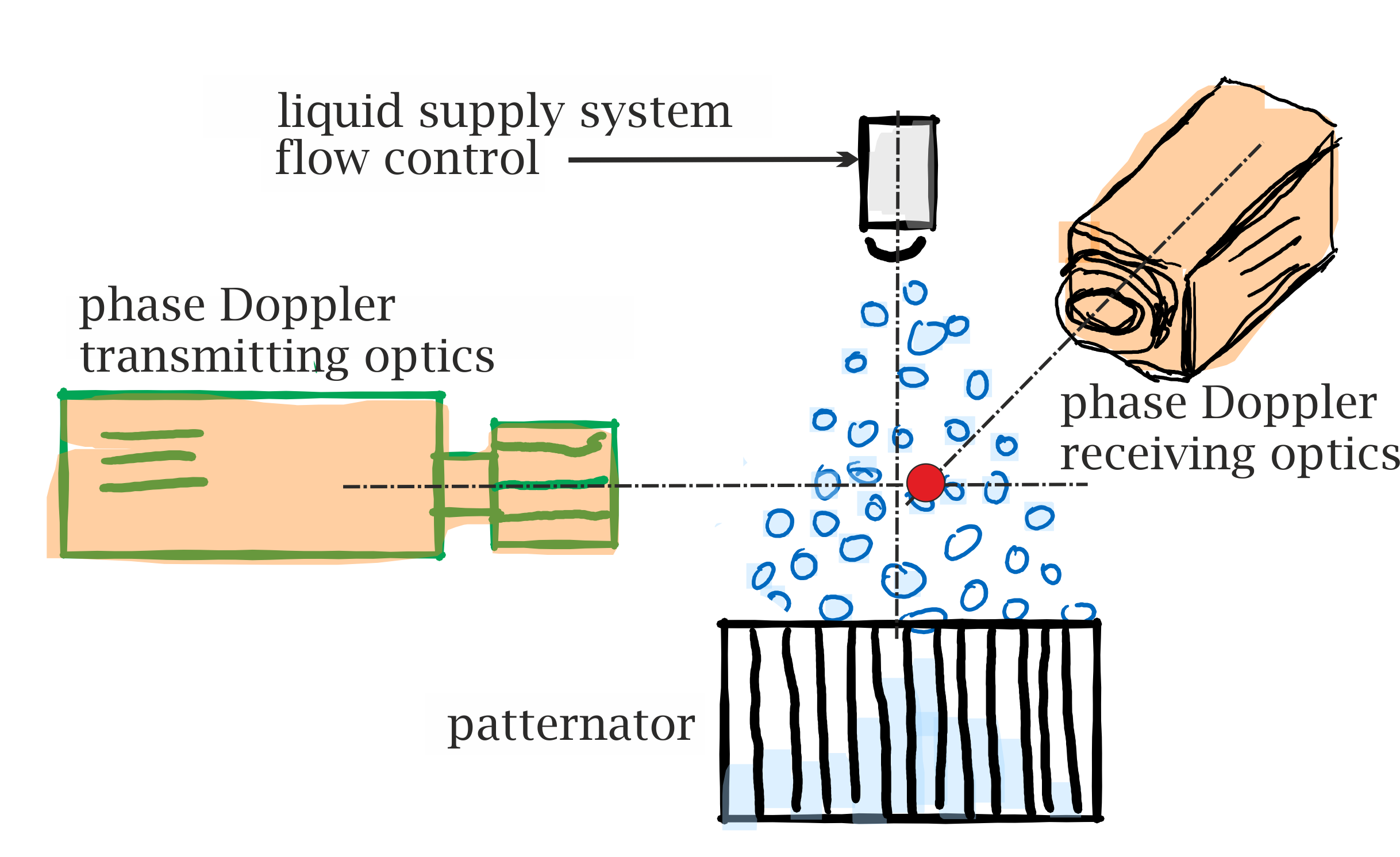}
    \caption{Sketch of the experimental setup used of the spray characterization.}\label{fig:expsetup}
\end{figure}

Various kinds of sprays were produced by conventional atomizers, which were accurately characterized using a phase Doppler system and a custom built patternator, as schematically shown in Fig.~\ref{fig:expsetup} right. The main  spray properties (mean diameter, mean velocity and mass flux) were determined for a wide range of the  operation conditions of the atomizer. The phase Doppler measurements were performed without the target, but at positions corresponding to specific locations immediately above the target; hence the spray parameters were local values.

A high speed camera equipped with a long distance microscope allows  observation of the hydrodynamic phenomena at the target surface  during spray impact.

The local heat flux and instantaneous surface temperature are derived from the measured temperature data using the computational solution of the inverse heat conduction problem \cite{Woodfield2006}. 

\begin{table*}[t]
	\caption{Overview the thermal properties of the materials and the value of the Leidenfrost temperature.}
	\centering
	\begin{tabular}{ p{1.9cm}  p{2cm} p{2cm}  p{4.3cm}  p{2cm}  p{1.5cm}} \hline
		Study & Type of \newline experiment & Fluid & Substrate material & $e_\mathrm{w} \, \newline \mathrm{(W s^{1/2}/m^2 K)}$ & $T_\mathrm{L} \, \mathrm{(^\circ C)}$ \\  \hline
		This study & Spray & Water & Stainless steel 1.4841  & $8.8501 \times 10^3$ & 342 \\ 
		This study & Spray & Water & Nickel Alloy 201  & $1.7892 \times 10^4$ & 286 \\ 
		\cite{Hoogendoorn1974} & Spray & Water & Stainless steel 1.4301  & $9.0193 \times 10^3$ & 371 \\ 
		\cite{Shoji1984} & Spray & Water & Nickel & $1.8569 \times 10^4$ & 321 \\ 
		\cite{Ito1991} & Spray & Water & Copper & $3.6476 \times 10^4$ & 134 \\ 
		\cite{Yao1987} & Spray & Water & Copper with chrome plating & $1.7831 \times 10^4$ & 222 \\ 
		\cite{Choi1987}  & Spray & Water & Copper with chrome plating & $1.7831 \times 10^4$ & 253 \\ 
		\cite{Yao2002} & Spray & Water & Copper with chrome plating & $1.7831 \times 10^4$ & 266 \\ 
		\cite{Bernardin1997b} & Drop chain & Water & Copper with gold plating & $2.7701 \times 10^4$ & 225 \\ 
		\cite{Tran2012} & Drop & Water & Silicon wafer & $9.7544 \times 10^3$ & 480 \\ 
		\cite{Wang2019a} & Drop & Water & FeCrAl & $6.5664 \times 10^3$ & 445 \\ 
		\cite{Wang2019a} & Drop & Water & Sintered SiC & $1.5733 \times 10^4$ & 350 \\ 
		\cite{Wang2019a} & Drop & Water & Zr-4 & $5.1511 \times 10^3$ & 531 \\ 
		This study & Sessile drop & Water & Copper & $3.6476 \times 10^4$ & 124 \\ 
		\cite{Baumeister1973} & Sessile drop & Water & Pyrex & $2.1030 \times 10^3$ & 700 \\ 
		\cite{Baumeister1973} & Sessile drop & Water & Brass & $1.9804 \times 10^4$ & 233 \\ 
		\cite{Baumeister1973} & Sessile drop & Water & Gold & $2.8174 \times 10^4$ & 184 \\ 
		\cite{Baumeister1973} & Sessile drop & Water & Aluminium & $2.0594 \times 10^4$ & 210 \\ 
		\cite{Bernardin1999} & Sessile drop & Water & Aluminium & $2.0594 \times 10^4$ & 175 \\ 
		\cite{Emmerson1975} & Sessile drop & Water & Monel & $9.5390 \times 10^3$ & 316 \\ 
		\cite{Bernardin1999} & Sessile drop & Water & Silver & $3.2925 \times 10^4$ & 157 \\ 
		\cite{Bernardin1999} & Sessile drop & Water & Graphite & $1.1690 \times 10^4$ & 310 \\ 
		\cite{Wachters1966} & Sessile drop & Water & Gold & $2.8174 \times 10^4$ & 170 \\ 
		\cite{Wang2019a} & Sessile drop & Water & FeCrAl & $6.5664 \times 10^3$ & 379 \\ 
		\cite{Wang2019a} & Sessile drop & Water & Sintered SiC & $1.5733 \times 10^4$ & 310 \\ 
		\cite{Wang2019a} & Sessile drop & Water & Zr-4 & $5.1511 \times 10^3$ & 415 \\ 
		\cite{Baumeister1973} & Sessile drop & Ethanol & Pyrex & $2.1030 \times 10^3$ &  360 \\ 
		\cite{Baumeister1973} & Sessile drop & Ethanol & Aluminium & $2.0594 \times 10^4$ & 155 \\ 
		\cite{Baumeister1973} & Sessile drop & Ethanol & Stainless steel & $9.0193 \times 10^3$ & 190 \\ 
		\cite{Bernardin1999} & Sessile drop & Acetone & Aluminium & $2.0594 \times 10^4$ & 134 \\ 
		\cite{Bernardin1999} & Sessile drop & Benzene & Aluminium & $2.0594 \times 10^4$ & 175 \\ 
		\cite{Bernardin1999} & Sessile drop & FC-72 & Aluminium & $2.0594 \times 10^4$ & 90 \\ 
		\hline	
	\end{tabular}
	\label{tab:TL-mat}
\end{table*}

\section{The  substrate materials and their thermal properties} 

For data taken from the literature, the  substrate materials and their thermal properties, type of experiment and the resulting Leidenfrost temperature are summarized in Table~\ref{tab:TL-mat}. The thermal properties of the designated substrate materials  are taken from general literature and handbooks. Since these properties are temperature dependent, they  have been calculated  at  the corresponding Leidenfrost temperature. In all cited studies the surface is polished. The liquid is water at ambient temperature. Since the spray or drop impact occurs at the surface, the thermal properties of the plating material is used in the case of a plated substrate, \citep{Yao1987, Choi1987, Yao2002, Bernardin1997b}. Data from \cite{Wang2019a} were  measured for different Weber numbers: those with a large $We$ were treated as spray data, whereas those with a small $We$ is treated as sessile drops.

\FloatBarrier
\bibliographystyle{apsrev4-2}
\bibliography{library}

\end{document}